\documentclass[aps,prb,groupedaddress,twocolumn,10pt]{revtex4-2}

\begin{document}

\newcommand{\be}{\begin{equation}}
\newcommand{\ee}{\end{equation}}
\newcommand{\bea}{\begin{eqnarray}}
\newcommand{\eea}{\end{eqnarray}}

\title{Diagnostic Tomography of Applied Holography}

\author{D. V. Khveshchenko} 
\affiliation{Department of Physics and Astronomy, 
University of North Carolina, Chapel Hill, NC 27599}

\begin{abstract}
\noindent
The single-particle behavior in 
$d\geq 1$-dimensional Fermi gases with a large number $N$ of species and strong 
short-range $s$-wave scattering is discussed in the $2d$
'tomographic' framework of a (pseudo)holographic correspondence with a certain $3d$ gravity   
of the $AdS_3$ type. However, due to the intrinsically topological nature 
of such a bulk theory its dynamics reduces to a purely boundary one and so,  
akin to its $SYK/AdS_2$ counterpart, this formal correspondence
neither represents a genuine case of, nor endorses the hypothetical 
generalized holographic duality. 
\end{abstract}

\maketitle

{\it Holography forever (yes/no?)}\\

After having had been abundantly present in the information space 
of strongly correlated systems for over $15$ years, the so-called 
'applied holographic' - a.k.a. 'bottom-up' $AdS/CMT$ (where CMT stands for Condensed Matter Theory but, in reality, often means $non-AdS/non-CFT$) - approach \cite{hol} 
has been finally disappearing under the radar, as of lately.
Once presented as spectacular (including quantitative, despite typically operating well outside the regime of 
applicability of any semiclassical approximation in the bulk gravity theory) 
successes in 'explaining' such popular condensed matter 
systems as the superconducting cuprates in their normal state, the bold claims of the early holographic studies are neither being elaborated upon even by the enthusiastic proponents, nor any longer scrutinized by the skeptics alike. 
In fact, nowadays even the remaining staunch advocates for applied holography  
tend to report on those physical phenomena which this approach 
$does$ $not$ - as opposed to those which is $does$ - capture \cite{bag}.

At its inception, though, this novel approach offered a seemingly straightforward and 
tempting to faithfully follow - 'no questions asked' - practical recipe 
which strove to provide a universal tool for studying a broad variety of
strongly-correlated quantum many-body systems \cite{hol}.
And under its self-proclamation of being uniquely suited for those situations
where the conventional condensed matter techniques were claimed to fail 
it has been opportunistically entertained in great many 
ways, thus creating a rather peculiar culture of prioritizing technical convenience over (self)critical 
judgment. Most frustratingly, the answer to the
central question: 'If generalized ($non-AdS/non-CFT$) holography were indeed valid, then $why$
would that be?' has neither been seriously demanded, nor pursued with any determination. 

Although somewhat delayed, the eventual demise of such a shortcut to bypassing the usual burden of proof, of course, should have been just a matter of time. However, it has also been suggested that, rather ironically, some of the general holographic claims might turn out 'to be right, albeit for the wrong reason' \cite{magic}.

For one, much of the late holographic discourse has been limited to the far more humble 
(and, incidentally, much less prone to criticism) topic as linear hydrodynamics,
one of the central issues being a possible 
(non)existence of fundamental bounds for the various kinetic coefficients 
and (non)universal relations between them \cite{bound}. 
However, this activity, too, has been gradually withering out, also due to the realization
that such relations readily arise within the standard theory of quantum transport  
and with no reference to any holographic conjecture whatsoever \cite{nussinov}. 

That said, it is quite conceivable that the truly systematic - as opposed to any heuristic - approach to generic many-body quantum systems could indeed be developed in the form of a path integral over some  Wigner function-type collective field variable acting in the corresponding phase-space. 
A subsequent reduction of this formally exact description 
to the first few moments of the Wigner's function, alongside their (non-linear) 
(hydro)dynamics, can then be thought of as establishing a holography-like relationship between the $2d+1$-dimensional 'bulk' and its boundary defined in some 
lesser ($d+1$, as the lowest) number of dimensions \cite{hydro}. 

In practice, such a (pseudo)holographic picture is likely to have much in common with the so-called 'geometric bosonization' technique that was discussed three decades ago \cite{bos}
and was recently resurrected (either with \cite{son} or without \cite{else} 
any reference to Ref.\cite{bos}).  
This prospective direction is still awaiting for its further development, though. 

Another holographic spin-off topic is signified by the 
Sachdev-Ye-Kitaev (SYK) model \cite{syk1}  
and related (random or not \cite{gurau}) $0+1$-dimensional models
which are often portrayed as the water-proof example of holographic duality. Curiously enough, this characterization would typically be made despite readily 
acknowledging that in $1+1$ dimensions  the 
pertinent Jackiw-Teitelboim (JT) bulk gravity appears to be non-dynamical
and is fully described by the boundary degrees of freedom, 
thereby revealing its intrinsically topological and effectively $0+1$-dimensional nature. 

Apart from some subtle high-energy features 
which require an introduction of extra bulk non-gravitational (matter) fields \cite{syk2}, 
the $0+1$-dimensional boundary dynamics determines all the long-distance bulk correlations, 
low-temperature thermodynamics, etc.
Moreover, this correspondence turns out to be 
very non-specific, as the dual $AdS_2$ geometry emerges universally
in the near-extremal regime of a generic $d+1$-dimensional bulk gravity \cite{hol}. 

Thus, although in the above cases certain many-body properties can indeed look like
being ostensibly holographic, such correspondence appears to represent 
a mere equivalence between the systems of, effectively, same dimensionality. 
And while still being potentially useful, such topological - akin to that in the 
Quantum Hall Effect - duality   
does not quite rise to the same level as genuine, 'non-topological', one
where the two dual systems do, in fact, belong in different dimensions.
In that regard, the former may be viewed merely as the result of using a redundant, non-minimal, description,
while the latter would indeed be radically new and requiring a major paradigm shift. 

The present note aims to add to the list of pseudo-holographic correspondences 
by suggesting that a certain class of Fermi systems 
with short-range, yet potentially strong, interactions (e.g., neutral Fermi gases) 
could be formally mapped onto the $2+1$-dimensional gravity of the $AdS_3$ variety.
However, the latter happens to be topological as well \cite{ads3}, 
and, therefore, is neither specific enough to positively identify the 
boundary system in question, nor practically
useful to allow for its properties to be accessible only via the bulk.\\  

{\it Tomographic fermions}\\

The physics of weakly dispersing bands,  
where correlation are strong and interactions can possibly dominate over
kinetic energy, has been the topic of many theoretical and experimental studies, 
starting with the Hubbard model and the Fractional
Quantum Hall effect, and culminating, in the recent years, with the
discovery of flat bands in twisted bi/tri-layer graphene and other systems.

It appears, however, that this 
fundamental problem finds its cleaner implementation in cold neutral Fermi gases. 
Unlike the former where the relevant interactions are often described by fixed  
intermediate (rather than truly strong) couplings under the conditions of  
broken translational and rotational invariances, the latter can 
be precisely tuned into the extreme (unitary) regime 
while maintaining translation and rotational invariance.

By allowing for the long-ranged Coulomb forces this model can be 
further applied to semiconductor heterostructures.
However, even in its most minimalistic formulation the full solution to this 
general problem still remains to be constructed. 
Arguably, it should have been the first target for applying 
any technique pretending to provide 
a new insight into such more complex scenarios as, e.g., 
the 'strange-metallic' normal state of the superconducting cuprates.
  
Lately, much attention has been paid to the random matrix-type models
whose hallmark feature is a dominant role of the so-called 'melonic' diagrams \cite{syk1,gurau,syk2}. 
In the spaceless SYK-type models, a justification of the melonic approximation requires
a large number $N$ of the fermionic flavors, both, in the random 
(original SYK) 
as well as the non-random (Witten-Gurau) (non)colored tensor models. 

Restoring spatial correlations and generalizing to the 
case of arbitrary $q\geq 4$-fermion couplings, 
the pertinent $d+1$-dimensional melonic theory is described
by the Luttinger-Ward functional 
\bea 
F=\sum^N_{a=1}\ln det(\partial_{\tau}-\epsilon_a({\bf \nabla}_{\bf x}))+\nonumber ~~~~~~~~~~~~\\
-
N\int_{\tau_1,{\bf x}_1}\int_{\tau_2,{\bf x}_2}
(\Sigma(\tau_{12},{{\bf x}_{12}} G(\tau_{12},{{\bf x}_{12}})-\nonumber~~~~~~~~~~\\
-{1\over q}
U(\tau_{12},{\bf x}_{12}) {G}^{q}({\bf x}_{12},\tau_{12}))~~~~~~~~~~~~
\eea
formulated in terms of the bi-local field variables $G$ and $\Sigma$ 
corresponding to the 
single-particle Green function
and its self-energy. The dispersion functions $\epsilon_a({\bf k})=v_a(k-k_{Fa})$
vanishing at the corresponding (Luttinger) Fermi surfaces (FS) endow the fermions 
with spatial dynamics while the coupling function 
$U(\tau,{\bf r})$ replaces the square of the SYK entangling amplitude 
averaged over a random ensemble (if any).
 
While the fine details of this function would be instrumental for
determining (non-universal) fermion structures at the lattice scale, 
the long-distance (potentially universal) properties can be fully 
characterized by the overall coupling strength and 
power-laws of its spatio-temporal decay.

The Schwinger-Dyson equations derived from Eq.(1) read 
\bea 
(\partial_{\tau_1}-\epsilon_a(-i{\bf \nabla_{\bf x1}}))G(\tau_{12},{{\bf x}_{12}})\nonumber\\
+
\int_{\tau_3,{\bf x}_{3}}{\Sigma}
(\tau_{13},{{\bf x}_{13}}){G}(\tau_{32},{{\bf x}_{32}})
=\delta({\bf x}_{12})\delta(\tau_{12})~~~~~~~~~~~
\eea
and
\be 
\Sigma(\tau,{\bf r})=U(\tau,{\bf r})G^{q/2}(\tau,{r})G^{q/2-1}(-\tau,-{r})
\ee  
which expression for 
the self-energy is formally similar to that of the $2^{nd}$ order Born approximation
in the Fermi gas with short-range (yet, potentially strong) scattering.

In the limit of a totally flat (degenerate) dispersion relations $\epsilon_a({\bf k})=const$ and 
strongly localized interaction, $U({\tau,{\bf r}})=J^2\delta({\bf r})$, the solution of Eq.(2) localizes in space 
\be
G_{syk-r}(\tau,{\bf r})\sim {sgn(\tau)\over (J\tau)^{\Delta}}\delta({\bf r})
\ee
with $\Delta=2/q$, as in the standard SYK model \cite{syk1}.
Correspondingly, its Fourier transform, $G_{syk-r}(\omega,{\bf k})\sim sgn(\tau)/J^{2/q}\omega^{1-2/q}$,
becomes independent of momentum. 
 
Introducing bare dispersions $\epsilon_a({\bf k})\neq const$ and/or a non-zero spatial 
range of interactions favors non-local solutions.  
For $q=4$ one such solution was proposed in Ref.\cite{patel}
where the use of Eqs.(2,3) was justified by  
introducing $N\gg 1$ radially nested FS, while 
the only allowed processes of fermion scattering would be resonant, 
the energies of the incoming and outgoing scattered particles obeying the resonant condition
$
\sum^{q/2}_{a=1}\epsilon_a({\bf k}_a)=\sum^{q}_{b=q/2+1}\epsilon_{b}({\bf k}_b)
$. 

A related version of this model was studied in Ref.\cite{loss}
by exploiting the 'tomographic' expansion of the propagator
\bea
G(\tau, {\bf r})
\approx
{k_F^{(d-1)/2}\over \Omega_r^{1/2}}
\sum_{\pm}e^{\pm i(k_Fr-\pi(d-1)/4)}
{g(\tau, \pm r)}
\eea
where $\Omega_r=S_{d-1}r^{(d-1)}$ is the surface area of the $d-1$-dimensional sphere and
$S_{d-1}=2\pi^{d/2}/\Gamma(d/2)$.

Expanding the $G$ factors in Eqs.(2,3) and discarding all but the $s$-wave amplitudes
(consistent with the assumed short-range nature of interactions),
one arrives at the equation for the $1+1$-dimensional propagator 
$g(\tau,r)$.
Although the radial variable $r$ takes values on the positive semi-axis,
the convergent and divergent spherical waves can be combined into
one chiral (uni-directional) mode defined in the entire unbounded domain $-\infty < r <\infty$, by analogy with the standard analysis of the Kondo effect. 
In contrast, for $d=1$ the spatial coordinate is initially unbounded and the 
fermion system is non-chiral. 

In the tomographic representation Eq.(2) then reduces to its 'radial' counterpart 
\bea 
(\partial_{\tau} +\partial_r)g(\tau,r)
+
\int_{\tau^{\prime},r^{\prime}}
\sigma(\tau-\tau^{\prime},r-r^{\prime})
g(\tau^{\prime},r^{\prime})~~~~~~~~~
\nonumber\\
=\delta(\tau)\delta(r)~~~~~~~~~~~~~~~
\eea
where the effectively $1+1$-dimensional self-energy reads
\be 
\sigma(\tau,{r})=
U(\tau,r)
\sum_{\xi_a,\zeta_a}e^{i\sum{k}_a{r}}
\prod_{a=2}^{q}{g(\eta_a\tau,\zeta_a{r})}
\Omega_r^{(4-q)/2}
\ee
where $\xi_a,\zeta_a=\pm 1$ are two-valued sign factors.
Among the different terms in the sum (7) the most important are the non-oscillating ones
with $\sum_a{k}_a=0$. Under such a selection rule only the terms with zero total $1d$ momentum
should be kept. Moreover, in the case of a common Fermi velocity the resonant scattering 
condition gets automatically fulfilled, too.

For the sake of generality it
would be instructive to consider the interaction function that decays algebraically
in, both, space and time, 
$
U(\tau,{r})={J^2/\tau^{2\alpha}{r}^{2\beta}}
$ 
with some positive $\alpha$ and $\beta$. 
In Eq.(7) it appears to be multiplied with 
the extra power ${(4-q)(d-1)/2}$ of $r$ which stems from the tomographic expansion (5).

For comparison, in Ref.\cite{loss}  the 'defect' power law governing the spatial dependence of
correlations was found to be $(2-q)(d-1)/2$, which mishap can be traced back 
to missing another factor of $\Omega_r$ which is due to the conversion 
of the $d$-dimensional spatial into the $1d$ radial $\delta$-function in the r.h.s. of Eq.(6). 

Neglecting the derivative terms in Eq.(6) gives rise to the generalized  SYK-like  integral equation
where the power-law decay of the interaction function prompts the use of factorized solution
\be 
g(\tau,{r})\sim {sgn\tau\over \tau^{2\Delta_{\tau}}}{1\over {r}^{2\Delta_{r}}}
\ee
which is dominated by the self-energy (7) and governed by 
the exponents 
\be 
\Delta_{\tau}={1-\alpha\over q}, ~~~~~~\Delta_{r}={1-\beta+(d-1)(1-q/4)\over q}
\ee
Such a solution was first proposed in the context of multi-dimensional SYK generalizations 
\cite{thick} and later reproduced in Ref.\cite{loss} 
(with no reference to Ref.\cite{thick} and, in fact,  
incorrectly, see above). Also, the factorized solution was 
utilized in a somewhat different - but mathematically similar - 
context of the replica-(a)symmetric SYK configurations \cite{maria}.

In the Fourier transform of the effectively $1d$ propagator 
\be 
g(\omega,{k})={1\over i\omega-\epsilon(k)-\sigma(\omega,k)} 
\ee
the bare (derivative) terms are combined with the 
Fourier-transformed self-energy 
\be 
\sigma(\omega,{k})={\lambda}\omega^{1-2\Delta_{\tau}}k^{1-2\Delta_r}
\ee
Balancing the bare against the self-energy 
terms allows one to estimate the effective dynamical critical index. 

Namely, in the complimentary regimes $\tau\gg r$ and $\tau\gg r$ one reads off the naive estimates
$z_{>}=(1-2\Delta_r)/2\Delta_{\tau}$
and
$z_{<}=2\Delta_{r}/(1-\Delta_{\tau})$
which, in general, would be inconsistent, as $z_>\geq 1$ while $z_<\leq 1$. 
In fact, for any $d$, $q$, and $\alpha,\beta>0$ 
the minimal attainable value of 
${z}_{>,min} = (1+\beta) /(1-\alpha) $ is always greater than the maximal ${z}_{<,max} = (1-\beta) /(1+\alpha)$. 

However, for $\alpha=\beta=0$ and $q=4$ (but arbitrary $d$) both estimates  
converge to the same value of $z=1$.
With this unique choice of parameters one identifies the regime 
where $\Delta_{\tau}=\Delta_{r}=1/4$ 
and the self-energy $\sigma(\omega,{k})={\lambda}(\omega k)^{1/2}$ governed by a dimensionless 
amplitude $\lambda\sim (J/\mu)^2$ with $\mu=v_Fk_F$ appears to be marginally on par with the bare terms 
in (10). 

These findings are consistent with the general 
expectation that, in contrast to the $0+1$-dimensional 
SYK/tensor models where all the $q$-particle couplings are 
strongly relevant in the infrared, in their (effectively) $1+1$-dimensional
generalizations the only marginal could be
the interactions with the lowest non-trivial value of $q=4$.

Incidentally, in this special case there exists an elegant 
exact solution to the integral equation (6) \cite{sondhi}  
\be
g(\tau,r)={1\over {\sqrt {(iv_+\tau-r)(iv_-\tau-r)}}}
\ee
where $v_{\pm}=v(1\pm \lambda)$. Eq.(12)   
is uniquely tailored to the case of $q=4$ and  
does not allow for any immediate generalization. 

Also, in the non-chiral case of $d=1$ the arguments presented in Ref.\cite{yunge}
suggest that the 'charge' mode tends to develop a gap due the generic back-scattering processes. 
Therefore, the following discussion will be limited to the dimensions $d>1$ where the FS is continuous
and the chiral tomographic fermions remain gapless.  

The Fourier transform of (12) 
\be
g(\omega,{k})={1\over {\sqrt {(i\omega-v_+k)(i\omega-v_-k)}}}
\ee
features the branch-cut singularity while exhibiting the intact dynamical critical exponent $z=1$. 
Instead of an anomalous power-law decay, though,
Eq.(13) demonstrates the behavior akin to spin-charge separation. 
Accordingly, the corresponding spectral function $Im g(\omega,k)$
is manifestly non-Fermi liquid (NFL)-like, 
showing a double-peak structure with the square-root singularities at $\omega=v_{\pm}k$.

The propagator given by Eqs.(5,12) differs markedly from the approximate (semi-numerical) solution \cite{patel} obtained 
with the use of the mixed $\tau-{k}$ representation 
in the momentum-space variant of the SYK model 
\be 
G_{syk-p}(\tau,{\bf k})\sim {sgn\tau\over (J\tau)^{1/2}}e^{-O(T/J)\epsilon(k)\tau} 
\ee
At low temperatures (13) is strongly localized in the coordinate space, regardless of its dimension, which behavior is suggestive of $z=\infty$ and the near-extremal 
$AdS_2\times R^d$ geometry of the corresponding bulk dual \cite{hol}. 

Likewise, Eq.(12) differs from the semi-local holographic propagator 
\be 
G_{semi-loc}(\omega,{\bf k})={1\over (i\omega)^{2\delta_k}-\epsilon(k)} 
\ee
where the momentum-dependent exponent $\delta_k=(a+b{\bf k}^2)^{1/2}$ 
varies continuously with the scaling dimension of the bulk fermion \cite{cubrovic}.
It has been used extensively in constructing the various 
holographic phenomenologies which, however, encounter significant problems with 
restoring the proper FS.\\

{\it Boundary-bound bulk dual}\\

Both, the $2d$- and $3d$-dimensional, gravities are intrinsically topological and, therefore, 
all their gauge invariant bulk degrees of freedom 
are effectively determined by the state of
the boundary \cite{bh}
Among other things, this implies that there are no gravitational waves
and the list of the classical background solutions is limited
to the maximally symmetric spaces of constant curvature: 
Minkowski, de-Sitter, or anti-de-Sitter. 

It was conjectured in Ref.\cite{hydro}, 
the pertinent class of quantum $1+1$-dimensional systems can be related to a formally 
$2+1$-dimensional phase-space geometry, the third ('holographic') 
coordinate then being played by the conjugate momentum $p$.

Somewhat surprisingly, there has been no consistent attempts 
to incorporate the $1d$ phenomenon of spin-charge separation into the holographic framework.
For one thing, the putative gravity dual corresponding to the 
$1+1$-dimensional chiral SYK was found in Ref.\cite{sondhi}
to attain the maximal $0+1$ SYK-like  
many-body quantum chaos quantified by the Lyapunov exponent
$\lambda_{L,max}=2\pi T$ right where the model loses its chiral nature ($v_-\to 0$).
 
Therefore, it is conceivable that a viable background geometry could be rid of the ubiquitous near-extremal black hole which exhibits 
the universal $AdS_2\times R^d$ metric 
as in its presence the boundary state would likely 
remain maximally chaotic for $T\to 0$ and arbitrary values of 
the entangling couplings \cite{syk1,syk2}.

As far as the task of identifying the pertinent geometry
is concerned, there exists a well-known connection \cite{ads3} between the $2+1$-dimensional 
gravity with a negative cosmological constant $\Lambda=-1/l^2$ 
\be
S={l\over 16\pi \kappa}\int d\tau drdp {\sqrt g}(R-\Lambda)
\ee
where $l$
and $\kappa$ are the AdS radius and Newton's constant, respectively, 
and the (double-sided) Chern-Simons model with the action 
\be
S={l\over 16\pi \kappa}
Tr\int d\tau dxdp \epsilon^{\mu\nu\lambda}({\hat A}_{\mu}^{\pm}\partial_{\nu} {\hat A}_{\lambda}^{\pm} + 
{\hat A}_{\mu}^{\pm}{\hat A}_{\nu}^{\pm}{\hat A}_{\lambda}^{\pm})
\ee 
Under the identification of the $3d$ metric 
\be
g_{\mu\nu}={l^2\over 4}(A_{\mu}^{+}-A_{\mu}^{-})(A_{\nu}^{+}-A_{\nu}^{-})
\ee
The equation of motion derived from (17) imposes the null curvature condition 
\be
\partial_{\mu}{\hat A}_{\nu}^{\pm}+{\hat A}_{\mu}^{\pm}{\hat A}_{\nu}^{\pm}-(\mu\leftrightarrow\nu)=0
\ee
which then becomes the Einstein equation, $R_{\mu\nu}-2g_{\mu\nu}\Lambda=0$ 
complemented by that of zero torsion.

The chiral connections ${\hat A}_{\mu}^{\pm}$ can be expanded  
in the basis of generators ${\hat L}^{\pm}_{0,\pm 1}$ 
of the algebra  
$SL(2,R)\times SL(2,R)=SO(2,2)$ which obey the commutation relations
$
[{\hat L}^{\pm}_n,{\hat L}^{\pm}_m]=(n-m){\hat L}^{\pm}_{n+m}
$
and are normalized as $Tr {\hat L}^{\pm}_{n}{\hat L}^{\pm}_m= {1\over 2}\delta_{n0}\delta_{m0}
-\delta_{n1}\delta_{m,-1}$.
Furthermore, the asymptotic symmetry algebra is, in fact, enhanced
to the product of two chiral Virasoro algebras.

Upon parameterizing the solutions of (18) in terms of 
an arbitrary group element ${\hat \chi}_{\pm}$,
functions $p_{\pm}(x_{\pm})$, and the conjugate chemical potentials
$\mu_{\pm}={\delta H^{\pm}\over \delta p_{\pm}}$  
\be
{\hat A}^{\pm}(x,p,\tau)={\hat \chi}^{-1}_{\pm}(p)
{\hat L}_0(\mu_{\pm}d\tau\pm p_{\pm}dx){\hat \chi}_{\pm}(p)
\ee
which becomes the continuity equation ${\dot p}_{\pm}\mp \mu_{\pm}^{\prime}=0$. 
By choosing the appropriate form of the Hamiltonian $H^{\pm}=\sum_la_lH^{\pm}_l$
where all $H^{\pm}_l=\sum_{k=1}^{l} b_k p_{\pm}^{k}\mu_{\pm}^{l-k}$ 
are in involution ($[H_k,H_l]=0$) one can then reproduce 
certain families of $1+1$-dimensional  
integrable (KdV, mKdV, Gardner, Benjamin-Ono, and other) equations \cite{bh}. 

On the gravity side, different saddle points of the coherent 
states path integral governed by (16)
can be identified 
as globally distinguishable (but locally $AdS_3$) classical solutions,
the two competing minima being the thermal $AdS_3$ 
and Banados-Teitelboim- Zanelli (BTZ)
black hole \cite{btz}.

In particular, the non-chiral spinless Luttinger liquid (LL)  
 with $k=1$ can be reproduced by introducing 
the original Brown-Henneaux boundary conditions
with constant $\mu_{\pm}\sim p_{\pm}$, the outer/inner BTZ horizons
being located at $p_{>/<}=(p_+\pm p_-)/2$. The dual metric 
\bea
ds^2={p^2dp^2\over (p^2-p^2_+)(p^2-p^2_-)}
+(p^2-p_{+}^2)({p^2-p^2_{-})\over p^2}d\tau^2~~~~~\nonumber\\
+p^2(dr-{p_{+}p_{-}\over p^2}d\tau)^2~~~~~~~~~~~~
\eea
describes a rotating BTZ black hole 
with the event horizon but no curvature singularity.
In particular, the non-rotating solution with $p_<=0$ can then be used to recover the left-right
symmetric LL (cf. the above discussion of the case $d=1$).

For static, yet non-constant, $p_{\pm}(x)$ the 
corresponding boundary solutions possess non-trivial global charges given by the 
chiral integrals of motion $H_{k}^{\pm}$ while their bulk counterparts   
can be regarded as black holes with multi-graviton excitations ('soft hair') \cite{bh}.
By introducing higher order terms $H^{\pm}_k$ with $k\geq 3$ 
one can generate new black hole configurations. 

The most general solution can be obtained by acting on
the ground state (e.g., BTZ black hole) with elements of the asymptotic symmetry
group commuting with the Hamiltonian. In this way, one can construct various constant curvature (locally $AdS_3$) space-times with anisotropic Lifshitz scaling and $z=2k-1$. 
This opens up the possibility of studying the systems with $z>1$ without 
the need of bulk geometries which are asymptotically Lifshitz space-times. 
Namely, in the case of a higher-spin symmetry $SL(M,R)$ with $M>2$ the list of 
attainable gravitational backgrounds includes the asymptotically 
Lobachevsky, Schroedinger, warped $AdS$, etc. space-times \cite{hs}.

Despite a whole range of possibilities it would be rather unlikely  
for the list of the (potentially) relevant gravity theories 
to include the Einstein-Maxwell-dilaton theory 
or for the relevant metrics to be even nearly as 
exotic as, e.g., Bianchi VII that has been repeatedly 
invoked in the holographic studies \cite{hol}.

In the context of the $1+1$-dimensional SYK model, 
however, the one-sided chirality with the causality cone falling within 
the range of velocities bounded by $v_{\pm}$ 
and the lack of conformal symmetry make it more difficult 
to identify the proper geometries.  

To that end, it was speculated in Ref.\cite{lian} that the properties of this model 
may be somewhat similar to the non-chiral CFTs with large central charges.
In particular, it was conjectured that their
gravity dual, if any, could be represented by a space-time rotating
faster than the speed of light at its boundary.
Also, the discrete (un)folding of the radial dimension performed  
in the process of constructing the 'tomographic' representation
might be suggestive of some orbifold-like underlying geometry. 
 
Some additional insight can be gained by comparing the boundary propagator (12) to the 
two-point function (in the Euclidean signature) of a bulk fermion of dimension 
$\Delta$. In the semiclassical regime it would be governed by the 
exponential of the action computed on the $3d$ geodesic ancored at the boundary end points
\be
{g}(\tau,r)\sim \exp(-\Delta\int{\sqrt {g_{pp}dp^2+g_{\tau\tau}d\tau^2+g_{rr}dr^2}}|_{p\to 0})
\ee
Fourier transforming this expression to the space-time domain one then obtains 
\be
g(\omega,k)\sim \exp(-\int dp
{\sqrt {g_{pp}({\Delta^2\over l^2}+{\omega^2\over g_{\tau\tau}}+{k^2\over g_{rr}}}})~~~)
\ee
Leaving a systematic identification of the pertinent metric to the future discussion,
by a direct inspection one finds that the propagator (12,13) can be reproduced
with the use of the metric 
\be
ds^2={dp^2\over p^2}+p^2(dr_-^2+\Delta v^2d\tau^2)
\ee
where $r_-=r-i\tau$ and the bulk fermion dimension $\Delta=1$. The logarithmic divergence 
in the momentum integral in (22,23) gets cut off at the low momenta 
$p\sim 1/{\sqrt {r_-^2+\delta v^2 \tau^2}}$,
thereby resulting in Eqs.(12,13). 

In the limit of vanishing interaction ( $\lambda=\delta v/2v\to 0$)  
the chiral boundary propagator and its Fourier transform return to their free counterparts  
$g_{0}(\tau,r)= 1/(r-i\tau)$ and  $g_{0}(\omega,k)=1/({i\omega-k})$, respectively.\\

{\it Strange theories of strange (or not so?) metals}\\

Since its inception, one of the most recurrent themes in applied holography was
the problem of 'strange metals', in reference to such electronic materials as the cuprates, pnictides,
heavy fermion compounds, twisted bi/tri-layer graphene, etc.

Thermodynamics of Eq.(1) 
is governed by the free energy ${F\sim} T^{1+1/z}$ in any dimension $d\geq 1$, 
thus producing a linear (possibly, $log$-enhanced) specific heat 
$C=-T\partial^2 F/\partial T^2 \sim T$ for $z=1$, 
albeit with a prefactor different from that in the ordinary Fermi gas.

For comparison, with the use of the solution (13) 
the longitudinal optical conductivity can be readily evaluated in the 
same `no vertex correction' approximation 
\bea
\sigma(\Omega)=
{\pi\mu\over 2T}
\int {d\epsilon_k d\omega\over \cosh^2(\omega/T)}
Im G(\omega+\Omega,{\bf {k}})Im G(\omega,{\bf {k}})
\nonumber\\
\sim {\lambda \mu T\over {(\lambda T)^2+\Omega^2}}~~~~~~~~~~~~ 
\eea
from which one recovers the Drude peak,  
$\sigma(\Omega)\sim \mu\delta(\Omega)$, in the vanishing coupling limit.

Albeit not immediately obvious, the source of momentum relaxation is provided by the intrinsic randomness built into the action (1). 
And while the customary substitution $\Omega\to T$ may not be generally justifiable 
(as it is in the 'no momentum drag' regime), the d.c. resistivity shows a linear 
behavior with increasing temperature, in agreement with the results of Ref.\cite{patel}. 

Furthermore, in Ref.\cite{patel} 
the slope of the linear in $T$ rate of momentum relaxation 
was found to be universal, which prediction appears to be consistent
with the data on a host of materials \cite{zaanen} (see, however, \cite{sad}).
Despite this intriguing observation, in the later work on the 
subject the remnant FS model was abandoned in favor of the more conventional 
coordinate space generalization of the 
SYK model with some additional single-particle randomness that, alongside
the random entangling correlations, yields the resistivity $\rho(T)=\rho_0+\alpha T$ \cite{esterlis}. 

The above results can be contrasted
against the earlier attempts to accommodate the general SYK scenario into the physics of cuprates and other 'strange metals' by exploiting the ultra-local 
behavior with $z=\infty$. This universality class was argued to be 
characterized by such hallmarks as universal ('Planckian') linear-$T$ equilibration/thermalization rate, maximal chaos, and saturated bounds for the kinetic coefficients \cite{zaanen}. 

Yet another thrust towards the multi-dimensional generalizations of the SYK model delivered the resistivity exponent $4/q$, thus suggesting some 'accidentally linear' temperature dependence for $q=4$ \cite{multi}. 
However, the general possibility of other values of this exponent should caution one against invoking this observation to explain the robust linear-$T$ dependence observed in the broad range of strongly correlated systems. 

Meanwhile, in the still other 'bottom up' holographic scenaria the thermodynamic and transport properties would be derived from such a popular workhorse 
as the Einstein-Maxwell-dilaton model which also paves the way for the Lifshitz and hyperscaling-violating (HV) background geometries which are parametrized
by the dynamical index $z$ and the HV exponent $\theta$. 
Besides, yet another exponent would often be introduced for the effective charge renormalization \cite{gut}.

The title of the aforementioned 
Ref.\cite{bag} states that the popular (Gubser-Rocha \cite{gr}) holographic model 
where the dynamical index $z$ and HV exponent $\theta$  
are both infinite while their 
ratio is finite, $\theta/z=-1$, does $not$ explain the cuprates
(presumably, suggesting that some alternate schemes might still do). 
Specifically, in Ref.\cite{bag} it was proposed to resolve 
the dichotomy between the $T$-dependences of resistivity and Hall angle observed in the cuprates by postulating a $T$-dependent  
carrier density $n\sim T$, alongside the generic relaxation time $\tau\sim 1/T^2$.   

Interestingly, though, despite including the earlier work of Ref.\cite{cuprates} 
in their list of references 
(item [36] in Ref.\cite{bag}) the authors of Ref.\cite{bag} did not seem to realize that in Ref.\cite{cuprates}  the same scenario had been proposed
(incidentally, the actual journal reference [36] in Ref.\cite{bag} was cited incorrectly). 

Nor, did the authors of Ref.\cite{bag} seem to appreciate that, apart from the conductivity 
$\sigma\sim 1/T$ and the Hall angle $\tan\theta_H\sim 1/T^2$, 
the scheme proposed in Ref.\cite{cuprates} 
renders the linear (possibly, $\log$- enhanced) specific heat and reproduces the experimentally observed magnetoresistivity, ${\Delta\rho\over \rho}\sim \theta^2_H\sim 1/T^{4}$, spin susceptibility (both, at the momentum $Q=(\pi,\pi)$ and momentum-integrated),
as well as the Hall Lorentz, Seebeck, and Nernst coefficients, 
the latter satisfying the relations
$
S\sim (T/e\sigma){d\sigma/d\mu}$ 
and 
$\nu_N\sim (T/eB){d\theta_H/d\mu}$ where $\mu$ is a chemical potential. 
Altogether, the above predictions appear to be in a generally 
better agreement with the data  
than the competing holographic phenomenologies \cite{hk} even without 
introducing an additional charge exponent.

Also, the proposal of Ref.\cite{cuprates}
differs from such historic scenaria as that of two distinct 
scattering times: $\tau_{tr}\sim 1/T$ and  $\tau_{H}\sim 1/T^2$ 
pertaining to the relaxation of longitudinal and transverse \cite{anderson}
or charge-symmetric vs anti-symmetric \cite{coleman} currents, respectively, or 
that based on the marginal Fermi liquid phenomenology \cite{varma}. 

Although the scenaria of $T$-dependent carrier density were indeed discussed at the early stages of the high-$T_c$ saga \cite{linear},  
the peculiar linear in $T$ behavior of the carrier density 
can be seen as problematic enough to view the above agreement 
with experiment as no more than fortuitous. 

Conceivably, however, in light of the gradually emerging consensus about the by and large mundane (hence, non-NFL) behavior in most of the cuprates phase diagram \cite{kitp1}
some of the seemingly unresolved issues might eventually turn out to 
be moot. On the other hand, such a conciliatory resolution could still be challenged farther down the road \cite{kitp2}.\\ 

{\it The bottom line (up front)}\\

Applied holography purportedly offers a unique insight into the largely uncharted territory of 'super-strongly coupled' systems with a nearly or even completely flat dispersion.
The majority of the recent and ongoing exercises on this topic exploits
the various variants of the 'ultra-local' scenario represented by the SYK model 
and its (pseudo)holographic dual which has been identified as the JT gravity. 
In contrast, the present note focuses on the holographic 
aspects (or lack thereof) of the intrinsically 'linear-radial' tomographic representation
models.  

On the practical side, a valid instance of holographic correspondence 
could then allow one to replace the conventional (usually impossible) 
exact or diagrammatic calculations in the strongly-interacting quantum 
boundary theory with the technically much simpler 
task of solving some classical equations $a$ $la$ Einstein in the higher-dimensional bulk. 

However, if the nature of the 
holography in question is topological, one can hardly gain any new knowledge 
by exploiting the (in this case, trivial) bulk gravity, thus largely 
negating any potential benefits of the alternate bulk description. In fact, 
all the essential properties of the SYK models can be understood without much 
input from the dual $AdS_2$ gravity, thus making their correspondence secondary to the  
(asymptotic) exact solubility of either model. 
The only 'holographic' aspect that appears to be essential for solving 
both models is the very existence of the 'conformal' solution (4),
reminiscent of the first such example 
found in the theory of phase transitions in helium \cite{helium}.

To summarize, in the present note it is argued that a broad family of $d\geq 1$-dimensional 
'tomographic' models allows for the unifying effective $1+1$-dimensional description which, in turn, can be reproduced in the context of some  bulk gravity of the $AdS_3$ type. 
Once again, the intrinsically topological nature of the latter 
theory does not provide any significant insight that could have not been 
gathered in the framework of the boundary model itself. 

On a more general note, 
the intrinsic topological aspect of any example of the $AdS/CMT$ holographic correspondence 
could be traced back to the popular interpretation of the extra 
holographic coordinate as that of the Wilson's renormalization group (RG) scale \cite{hol}. 
Despite being intuitively appealing, this identification may run  
into a potential problem with double counting of the 
effects of the boundary degrees of freedom which are supposed to generate that RG flow without any reference to the bulk fields. 
Indeed, any coarse-graining RG procedure (either in the Wilson's or in tensor network sense) can be executed solely within the boundary theory. Introducing an extra holographic coordinate would then be redundant and, therefore, subject to a gauge condition and/or an associated constraint. 
 
The above discussion implies that any purported discovery of a novel $AdS/CMT$ holographic duality should be calling for an inquiry into whether it really involves 
the systems in different dimensions ( 'It from Qubit', as per the popular motto \cite{hol}) 
or is it merely topological ('All from Hall', as per the above slogan's obscure rival \cite{hall}). Contrary to the former, the 'HALLographic' correspondence 
would be relating pairs of systems of the (de facto) same dimensionality and, therefore, would be unlikely to serve as a preferred means of gaining some 
exclusive insight into the properties of one party ('boundary')  
by instead studying its counterpart ('bulk'). 

The hospitality at the Aspen Center for Physics
funded by National Science Foundation through the grant PHY-2210452 
and at the International Institute of Physics in Natal, Brazil
supported by the Simons Foundation are gratefully acknowledged.

\end{document}